\documentclass[10pt,twoside]{mathdoc}

\usepackage{distoperators}

\usepackage{IEEEtrantools}

\title{Exact and Approximate Constants of Motion\\ in Stochastic Contact Processes}
\author{Dami\'an H. Zanette}
\author{Eric A. Rozan}
\affil{Centro At\'omico Bariloche and Instituto Balseiro, Comisi\'on Nacional de Energ\'{\i}a At\'omica, Universidad Nacional de Cuyo, Consejo Nacional de Investigaciones Cient\'{\i}ficas y T\'ecnicas, 8400 San Carlos de Bariloche, R\'{\i}o Negro, Argentina }

\makeatletter
\def\runauthor{D. H. Zanette and E. A. Rozan}
\def\runtitle{Exact and Approximate Constants of Motion in Stochastic Contact Processes}
\makeatother

\begin{document}
\maketitle
\thispagestyle{empty}
\begin{abstract}
    We study a variety of stochastic contact processes --directly related to models of rumor and disease spreading-- from the viewpoint of their constants of motion, either exact or approximated. Much as in deterministic systems, constants of motion in stochastic dynamics make it possible to reduce the number of relevant variables, confining the set of accessible states, and thus facilitating their analytical treatment. For processes of rumor propagation based on the Maki-Thompson model, we show how to construct exact constants of motion as linear combinations of conserved quantities in each elementary contact event, and how they relate to the constants of motion of the corresponding mean-field equations, which are obtained as the continuous-time, large-size limit of the stochastic process. For SIR epidemic models, both in homogeneous systems and on heterogeneous networks, we find that a similar procedure produces approximate constants of motion, whose average value is preserved along the evolution. We also give examples of exact and approximate constants of motion built as nonlinear combinations  of the relevant variables, whose expressions are suggested by their mean-field counterparts.
\end{abstract}

\section{Introduction}

Constants of motion are central to the analysis of dynamical processes because they constrain the regions of state space that a system can access, thereby simplifying its mathematical description \cite{Arnold,Strogatz}. If a physical system admits one or more invariants --be they conserved energies, masses, charges, or more abstract integral quantities-- its dynamics become confined to lower-dimensional manifolds within the original state space. This reduction is not merely computationally convenient, but provides structural insight into the qualitative behavior of the system. In deterministic dynamics, constants of motion can often be used to construct reduced models or to identify families of trajectories parameterized by invariant quantities. Even in systems that are not fully integrable, partial sets of invariants can help isolate slow manifolds, determine stability boundaries, and identify geometrical constraints governing long-term behavior \cite{Saletan}. Although mechanical systems supply the canonical examples of conservation laws --energy, linear and angular momentum-- this specific framework is only one instance of a far broader methodological scaffold. Noether's theorem \cite{Goldstein}, which links conserved mechanical quantities to symmetries, remains important as a conceptual foundation but, in many other contexts, constants of motion are not tied to continuous symmetries in any strict sense.

Beyond classical mechanics, constants of motion appear in a wide variety of domains where they play a similarly structural, though sometimes subtler, role. In stochastic processes, for example, conservation laws often take the form of invariant probability fluxes, conserved moments, or linear combinations of state variables that remain unchanged under the stochastic evolution \cite{vanKampen,Reichl}. These invariants can be crucial for deriving stationary distributions, identifying metastable sets, and reducing chemical reaction networks or population models to lower-dimensional effective descriptions. In chemical kinetics, conserved quantities restrict the dynamics of reaction networks to affine subspaces, allowing substantial dimensionality reductions even when the underlying reaction kinetics are nonlinear \cite{Haken,chem}. In the physics of fluids and kinetic theory, generalized Hamiltonian structures give rise to Casimir invariants, which constrain the admissible evolution of distribution functions and help classify equilibria and stability properties \cite{Morrison,Casimir}. Across all these settings, constants of motion act as organizing principles: they delimit the possible changes of the system along time, restrict its long-term behavior, and offer a unifying language for reduction, stability analysis, and qualitative classification.

In this contribution, we focus on the role of constants of motion in the analysis of stochastic contact processes. These are aleatory processes defined on systems formed by agents, where the state of each agent changes with time as a result of the interaction with other peers \cite{Liggett}. As referenced below, contact processes play a fundamental role in the mathematical modeling of a wide class of phenomena in biological and socioeconomic systems, covering disease dissemination, information transmission, and commercial or financial transactions, among many others. Here, for a variety of stochastic contact processes related to the modeling of rumor and disease propagation, we construct linear constants of motion and show how they help obtaining information on the evolution of the system. We demonstrate that, while some processes admit exact nontrivial constants of motion, others can only have approximated conserved quantities, that fluctuate randomly around their initial value as time elapses. In all cases, we compare with the conserved quantities known to exist in the mean-field version of each process. We also explore the possibility of finding nonlinear constants of motion.  

\section{Exact constants of motion in Maki-Thompson rumor propagation} \label{MTrp}

Stylized models of rumor propagation have become one of the main applications of contact processes in the mathematical approach to social phenomena. Starting from a handful of abstract stochastic processes proposed several decades ago \cite{Daley,MT,Sudbury}, these models have been extended to encompass a wide class of factors that influence information transfer in social systems, from  mass-media dissemination \cite{MM}, to networked structures \cite{net1,net2}, to population heterogeneity \cite{SciRep}. 

Among the most traditional models, the Maki-Thompson process \cite{MT} emulates the propagation of a rumor over a homogeneous population formed by $N$ agents. Evolution goes on by discrete steps. At any step, each agent can be in one of three states which, by analogy with epidemiological models (see Section \ref{SIR}), are named susceptible (S), infected (I), and refractory (or recovered, R). Susceptible agents have not heard the rumor yet, infected agents have heard the rumor and are willing to transmit it, and refractory agents have heard the rumor but have lost interest and do not transmit it. At each evolution step, one of the infected agents is selected at random and a second agent is chosen, also at random, from the remaining of the population. If the second agent is susceptible, it becomes infected, representing the transmission of the rumor. If, on the other hand, the second agent is infected or refractory, the first agent looses interest in the rumor and becomes refractory. Typically, the process is run starting from an initial condition with many susceptible agents (for instance, $S_0=N-1$), a few infected agents (for instance, $I_0=1$), and no refractory agents ($R_0=0$). Evolution ends when no more infected agents are left in the population. The first column of Table \ref{tab1} shows the two possible events in the format of chemical reactions. In the second column, we quote the corresponding probabilities of each reaction at step $n$, where $S_n$ is the number of susceptible agents at that step.

\begin{table}[H] 
\caption{Events in the Maki-Thompson model. First column: each event represented as a chemical reaction between susceptible (S), infected (I), and refractory (R) agents. In the second row, $\{ {\rm I},{\rm R} \}$ stands for either I or R. Second column: probability of each event. Third column: constants of motion associated with each state (S, I, and R) in each event.
\label{tab1}}
\begin{tabularx}{\textwidth}{lll}
\toprule
\textbf{event}	& \textbf{probability}	 &\textbf{constants of motion}  \\
\midrule
${\rm I}+{\rm S} \to {\rm I}+{\rm I}$ & $S_n/(N-1)$  & $S_n+n$, $I_n-n$, $R_n$ \\
${\rm I}+\{ {\rm I},{\rm R} \} \to {\rm R}+\{ {\rm I},{\rm R} \}$ &  $1-S_n/(N-1)$ &  $S_n$, $I_n+n$, $R_n-n$\\
\bottomrule
\end{tabularx}
\end{table}

\subsection{Constants of motion} \label{CoM1}

For each event, we have  constants of motion  associated with each state S, I, and R. Since in a transmission event (first row of Table \ref{tab1}), a susceptible agent becomes infected, the number of susceptible agents, $S_n$, plus the index of the evolution step, $n$, remains constant. Likewise, the difference $I_n-n$ is a constant of motion. Finally, since the number of refractory agents does not change, $R_n$ remains constant as well. These three constants of motion are shown in the third column of the table. Analogous arguments show that, for an event where an infected agent becomes refractory (second row of Table \ref{tab1}), the associated constants of motion are $S_n$, $I_n+n$, and $R_n-n$. Naturally, moreover, any combination of the three constants of motion is also a constant of motion for each event. 

Can we combine the constants of motion quoted above in such a way as to obtain a function of $S_n$, $I_n$, $R_n$, and $n$ that is a conserved quantity for {\em both} events? To answer this question, we consider linear combinations of  the constants of motion, requiring that the identity
\begin{equation} \label{req1}
    \alpha_1 (S_n+n)+\beta_1(I_n-n)+\gamma_1 R_n = \alpha_2 S_n+\beta_2(I_n+n)+\gamma_2 (R_n-n) 
\end{equation}
holds for a suitable choice of the coefficients. In order to get a solution with constant coefficients, we must have $\alpha_1=\alpha_2\equiv \alpha$, $\beta_1=\beta_2\equiv \beta$, and $\gamma_1=\gamma_2\equiv \gamma$. Equation (\ref{req1}) then implies
\begin{equation} \label{cond1}
    \alpha-2\beta+\gamma = 0.
\end{equation}
This linear equation has two independent solutions, for instance, $\alpha=\beta=\gamma=1$, and $\alpha=2$, $\beta=1$, $\gamma=0$. Replacing in the combinations of Eq.~(\ref{req1}), we get
\begin{eqnarray}
    K_1&=&S_n+I_n+R_n, \label{K1} \\
    K_2&=&2S_n+I_n+n, \label{K2}
\end{eqnarray}
which are two quantities conserved in both events. In the first of these joint constants of motion, we recognize the total number of agents, $K_1=N$, which is trivially conserved due to the fact that no agent is created or destroyed in any of the events. On the other hand, $K_2$ is a nontrivial conserved quantity, combining the number of susceptible and infected agents, and the index of the evolution step  \cite{Duan}. Its value depends on the specific initial condition chosen to run the process: $K_2=2S_0+I_0$. In the following, we show how $K_2$ helps to get information on the evolution.      

\subsection{Solution for the mean number of agents in each state} \label{MN1}

Since the probability of each event (second column of Table \ref{tab1}) depends on $S_n$ only, it is possible to write a master equation for the probability that the number of susceptible agents is $S$ as step $n$, $P_n(S)$, disregarding the number of infected and refractory agents. It reads
\begin{equation} \label{ME1}
    P_{n+1} (S) = \frac{S+1}{N-1} P_n(S+1)+\left(1-\frac{S}{N-1} \right) P_n(S).
\end{equation}
From this master equation, we immediately find the evolution law for the mean number of susceptible agents, $\langle S \rangle_n = \sum_{S=0}^N S P_n(S)$,
\begin{equation} \label{deltaS}
    \langle S \rangle_{n+1} = \left( 1-\frac{1}{N-1}\right)\langle S \rangle_n ,
\end{equation}
whose solution reads
\begin{equation} \label{Sn1}
    \langle S \rangle_n = \left( 1-\frac{1}{N-1}\right)^n S_0.
\end{equation}
Here, $\langle \cdot \rangle_n$ denotes the mean value, calculated at step $n$, over realizations of the stochastic process starting from the same initial condition $S_0$.  Averaging Eq.~(\ref{K2}) over realizations with the same initial condition and using Eq.~(\ref{Sn1}), we get the mean number of infected agents:
\begin{equation} \label{In1}
     \langle I \rangle_n =I_0+ 2S_0 \left[ 1- \left( 1-\frac{1}{N-1}\right)^n\right]-n.
\end{equation}
The mean number of recovered agents is just $\langle R \rangle_n=N- \langle S \rangle_n- \langle I \rangle_n$. 

The solution for $\langle I \rangle_n$, Eq.~(\ref{In1}), makes it possible to compute, although not always analytically, some important average features of the stochastic process. For instance, in the limit of large $N$, where $[1-(N-1)^{-1}]^n \approx \exp(-n/N)$, we find the number of steps at which the number of infected agents reaches a maximum, and the value of $\langle I \rangle_n$ at the maximum, respectively, 
\begin{equation}
n_{\rm max} = N \ln \frac{2S_0}{N}, \ \ \ \ \ 
\langle I \rangle_{\rm max} = I_0+2S_0 - N \left( 1+ \ln \frac{2S_0}{N}\right).
\end{equation}
Note that these results make sense if $S_0> N/2$. Otherwise, the number of infected agents does not reach a maximum during the evolution. By equating $\langle I \rangle_n=0$, meanwhile, we can find implicit equations for the expected evolution step at which the process ends, and for the corresponding average number of susceptible agents. In the limit of large $N$, we respectively have 
\begin{equation}
    n_{\rm end} = I_0 +2S_0 \left[ 1-\exp \left(-\frac{n_{\rm end}}{N}\right)\right], \ \ \ \ \ 
\langle S \rangle_{\rm end} = S_0 \exp \left(-\frac{n_{\rm end}}{N}\right) .
\end{equation}
For the initial condition $S_0\to N$ and $I_0\to 0$, we find $n_{\rm end} \approx 1.594 N$, and obtain the classical result for ``the proportion of the population never hearing a rumor'' \cite{Sudbury,Duan}, $\langle S \rangle_{\rm end}=0.203 N$.

\subsection{Mean-field evolution}  \label{MF1}

Mean-field differential equations for the Maki-Thompson model are found in the limits of a large number of agents and continuous time. Firstly, we note that, in continuous time, the sequential choice of infected agents of the stochastic model must be replaced by a scheme where, on the average, each infected agent has the opportunity of interacting with the rest of the population once per time unit. This is achieved if the duration of each evolution step is proportional to the inverse of the number of infected agents, concretely,
\begin{equation} \label{tau}
    t_{n+1}-t_n =\frac{\tau}{I_n},
\end{equation}
where $t_n$ is the time of occurrence of step $n$, $I_n$ is the corresponding number of infected agents, and $\tau$ is a constant that fixes time units. Next, combining Eqs.~(\ref{deltaS}) and (\ref{tau}), we estimate the time derivative of the number of susceptible agents as
\begin{equation}
   \frac{\text d S}{\text d t} \approx \left\langle \frac{1}{ t_{n+1}-t_n}\right\rangle \left(\langle S \rangle_{n+1}-\langle S \rangle_n \right) = -\frac{1}{\tau N}\langle S \rangle_n \langle I \rangle_n.
\end{equation}
Within this approximation, we get
\begin{eqnarray} \label{MTMF1}
    \frac{\text d s}{\text d t}= -\frac{si}{\tau},
\end{eqnarray}
where $s(t)=\langle S \rangle_n /N$ and $i(t)=\langle I \rangle_n /N$ are the average fractions of susceptible and infected agents, respectively. Proceeding in the same way for the time derivatives of the number of infected and refractory agents, we find
\begin{equation}  \label{MTMF2}
     \frac{\text d i}{\text d t}=  \frac{si -i(i+r)}{\tau}, \ \ \ \ \ \frac{\text d r}{\text d t}= \frac{i(i+r)}{\tau},
\end{equation}
with $r(t)=\langle R \rangle_n /N$.

Equations (\ref{MTMF1}) and (\ref{MTMF2}) provide the mean-field description of the Maki-Thompson model. Inspection immediately shows that the sum $s+i+r$ is a constant of motion. This represents, of course, the conservation of the total number of agents, thus corresponding to the constant of motion of Eq.~(\ref{K1}). From the definition of the fractions $s$, $i$, and $r$, we have $s+i+r=1$. 
A second constant of motion can be obtained by calculating the ratio between the first of Eqs.~(\ref{MTMF2}) and Eq.~(\ref{MTMF1}):
\begin{equation}
    \frac{\text d i}{\text d s}=- \frac{si-i(i+r)}{si}=-2+\frac{1}{s},
\end{equation}
where we have replaced $r=1-s-i$. Integrating with respect to $s$ from reference values $s_0$ and $i_0$, we find the second constant for the mean-field equations:
\begin{equation}
    K_{\rm mf} = 2s+i-\ln \frac{s}{s_0} \equiv 2 s_0+i_0.
\end{equation}
Remarkably, although the functional form of this conserved quantity does not coincide with that of $K_2$ in Eq.~(\ref{K2}), their numerical values in the mean-field limit are the same (up to a factor $N$): $K_2=NK_{\rm mf}$. In fact, identifying the reference values $s_0$ and $i_0$ with the initial condition,  $s_0=S_0/N$ and $i_0=I_0/N$, we have 
\begin{equation} \label{equiv}
    \ln \frac{s(t)}{s_0} = \ln \frac{\langle S \rangle_n}{S_0} = -\frac{n}{N} ,
\end{equation}
where we have used Eq.~(\ref{Sn1}) in the limit of large $N$. The two constants of motion of the mean-field equations play the same role as $K_1$ and $K_2$ in the Maki-Thompson model, and can be used to get the mean-field counterpart of the results found in Section \ref{MN1} for the stochastic process.

\subsection{Two extensions of the Maki-Thompson model}

In this section, we analyze two extensions of the Maki-Thompson stochastic process, which complexify the model by adding more states accessible to agents, but can be treated along the same lines as in the preceding sections. The first extension introduces resistance to recovery for infected agents \cite{Duan}. Specifically, an infected agent must attempt transmission to infected or refractory agents a total of $M$ times before becoming in turn refractory. This variant is formulated by introducing $M$ states of infection, ${\rm I}^{(m)}$ ($m=0,\dots , M-1$), where $m$ indicates the number of transmission events already attempted by the agents in that state. Table \ref{tab2} summarizes the relevant events, probabilities, and constants of motion. The original model is recovered for $M=1$.

\begin{table}[H] 
\caption{Events in the first extension of the Maki-Thompson model, where $M$ transmission attempts to infected or refractory agents are needed for an infected agent to become refractory. In the first column, I indicates any of the I$^{(m)}$. In the second column, $I_n=\sum_{m=0}^{M-1} I^{(m)}_n$. For brevity, the third column only shows the constants of motion which combine numbers of agents and the index of the evolution step $n$. All the other numbers of agents are also constants of motion. The first and second rows hold for $m=0,\dots, M-1$ and $m=0,\dots, M-2$, respectively. In the second and third rows, $\{ {\rm I},{\rm R} \}$ stands for either I or R.\label{tab2}}
\begin{tabularx}{\textwidth}{lll}
\toprule
\textbf{event}	& \textbf{probability}	 &\textbf{ constants of motion}\\
\midrule
${\rm I}^{(m)}+{\rm S} \to {\rm I}^{(m)}+{\rm I}^{(0)}$ & $I^{(m)}_n S_n/I_n(N-1)$  & $S_n+n$,  $I^{(0)}_n-n$    \\
${\rm I}^{(m)}+\{ {\rm I},{\rm R} \} \to {\rm I}^{(m+1)}+\{ {\rm I},{\rm R} \}$ &  $I^{(m)}_n[1-S_n/(N-1)]/I_n$  &    $I_n^{(m)}+n$, $I_n^{(m+1)}-n$   \\
${\rm I}^{(M-1)}+\{ {\rm I},{\rm R} \} \to {\rm R}+\{ {\rm I},{\rm R} \}$ &  $I^{(M-1)}_n[1-S_n/(N-1)]/I_n$  & $I_n^{(M-1)}+n$,  $R_n-n$\\
\bottomrule
\end{tabularx}
\end{table}

Proceeding as in Section \ref{CoM1}, we can find $M$ equations for the $M+2$ coefficients of a linear combination which gives a constant of motion for all the events of the model. We thus have two independent solutions. Not unexpectedly, one of them corresponds to the trivial constant of motion $N=S_n+\sum_{m=0}^{M-1} I^{(m)}_n+R_n$, namely, the total number of agents. The second, nontrivial joint constant of motion is 
\begin{equation} \label{KMT22}
    K=(M+1) S_n +\sum_{m=0}^{M-1} (M-m) I^{(m)}_n+n .
\end{equation}
For this extension of the Maki-Thompson model, the probability of having $S$ susceptible agents at step $n$, $P_n(S)$, also satisfies the master equation (\ref{ME1}), so that the solution for $\langle S \rangle_n$ given by Eq.~(\ref{Sn1}) still holds. Now, however, the existence of a constant of motion cannot be used to find the number of infected agents $I^{(m)}$ for each value of $m$, but rather the combination $\sum_{m=0}^{M-1} (M-m) I^{(m)}_n$. 

The mean-field equations for the present model are
\begin{equation}
    \begin{split}
        \frac{\text d s}{\text dt } &= -\frac{s}{\tau} \sum_{m=0}^{M-1} i^{(m)} ,\\
        \frac{\text d i}{\text dt }^{(0)} &= \frac{s}{\tau} \sum_{m=0}^{M-1} i^{(m)}-\frac{1-s}{\tau} i^{(0)} ,  \\
        \frac{\text d i}{\text dt }^{(m)} &= \frac{1-s}{\tau} \left[i^{(m-1)}- i^{(m)} \right], \ \ \ \ \ m=1,\dots, M-1, \\
        \frac{\text d r}{\text dt } &= \frac{1-s}{\tau}  i^{(M-1)}, \\
    \end{split}
\end{equation}
with $i^{(m)} = \langle I^{(m)} \rangle/N$, and $s$ and $r$ defined as in the preceding section. Direct calculation shows that, together with $s+\sum_{m=0}^{M-1} i^{(m)}+r=1$, the quantity
\begin{equation}
    K_{\rm mf}=(M+1) s+\sum_{m=0}^{M-1} (M-m) i^{(m)}- \ln \frac{s}{s_0}= (M+1) s_0+\sum_{m=0}^{M-1} (M-m) i_0^{(m)}, 
\end{equation}
with $s_0$ and $i_0^{(m)}$ the initial values of the corresponding variables. For large $N$, in full analogy with the original Maki-Thompson model, the numerical value of $K_{\rm mf}$ coincides with that of $K$ in Eq.~(\ref{KMT22}) up to a factor $N$: $K=N K_{\rm mf}$.

\vspace{10 pt}

In the second extension of the Maki-Thompson model considered here, which incorporates resistance to infection, a susceptible agent must be contacted $M$ times by any infected agent to become infected. To formulate this variant, we introduce $S_n^{(m)}$ as the number of susceptible agents, that, at step $n$, have already been contacted $m$ times by infected agents ($m=0,1,\dots, M-1$). Events, probabilities, and constants of motion are listed in Table \ref{tab3}. The original model is reobtained for $M=1$.

\begin{table}[H] 
\caption{As in Table \ref{tab2}, for the second extension of the Maki-Thompson model, where $M$ transmission attempts to a susceptible agent are needed for the agent to become infected.  In the second column, $S_n=\sum_{m=0}^{M-1} S^{(m)}_n$. The first row holds for $m=0,\dots, M-2$.\label{tab3}}
\begin{tabularx}{\textwidth}{lll}
\toprule
\textbf{event}	& \textbf{probability}	    & \textbf{constants of motion}  \\
\midrule
${\rm I}+{\rm S}^{(m)} \to {\rm I}+{\rm S}^{(m+1)}$ & $  S^{(m)}_n/(N-1)$  & $S_n^{(m)}+n$, $S_n^{(m+1)}-n$   \\
${\rm I}+{\rm S}^{(M-1)} \to {\rm I}+  {\rm I}$ &  $ S^{(M-1)}_n/(N-1)$  & $S^{(M-1)}_n+n$, $I-n$ \\
${\rm I}+\{ {\rm I},{\rm R} \} \to {\rm R}+\{ {\rm I},{\rm R} \}$ &  $1-S_n/(N-1)$ &  $I_n+n$, $R_n-n$\\
\bottomrule
\end{tabularx}
\end{table}

As in the first extension of the Maki-Thompson process, two independent constants of motion can be found. The first, trivial one is the total number of agents, $N=\sum_{m=0}^{M-1} S^{(m)}_n+I_n+R_n$. The second joint constant of motion reads
\begin{equation} \label{KMT23}
    K=\sum_{m=0}^{M-1} (M-m+1) S^{(m)}_n+I_n+n .
\end{equation}

Now, the master equations for the different sub-classes of susceptible agents are coupled to each other, and so are the equations for the corresponding average numbers of agents. Nevertheless, it is possible to find their general solution. For an initial condition where no susceptible agent has yet been contacted by an infected agent, $S^{(0)}_0 = S_0$ and $S^{(m)}_0 = 0$ for $m=1,\dots, M-1$, the solution reads
\begin{equation}
    \langle S^{(m)}\rangle_n = \binom{n}{m}\left(1-\frac{1}{N-1} \right)^{n-m} \frac{S_0}{(N-1)^m}.
\end{equation}
Replacing into Eq. (\ref{KMT23}) gives the average number of infected agents $\langle I\rangle_n$. With this result, it is possible to evaluate the step $n$ at which the number of agents reaches a maximum, the height of the maximum, the total duration of the process, and the number of surviving susceptible agents, much as done in Section \ref{MN1} for the original model. These calculations are not included here for the sake of brevity. 

In the mean-field limit we have
\begin{equation}
    \begin{split}
        \frac{\text d s}{\text dt }^{(0)} &= -\frac{s^{(0)} i}{\tau} , \\
        \frac{\text d s}{\text dt }^{(m)} &= \frac{i}{\tau} \left[s^{(m-1)}- s^{(m)} \right], \ \ \ \ \ m=1,\dots, M-1 ,\\
        \frac{\text d i}{\text dt } &= \frac{s^{(M-1)} i}{\tau} -\frac{1}{\tau} \left( 1-\sum_{m=0}^{M-1} s^{(m)} \right)i  ,  \\
        \frac{\text d r}{\text dt } &= \frac{1}{\tau} \left( 1-\sum_{m=0}^{M-1} s^{(m)} \right)i, \\
    \end{split}
\end{equation}
with $s^{(m)} = \langle S^{(m)} \rangle/N$, and $i$ and $r$ defined as in the preceding section. The nontrivial constant of motion for these equations is 
\begin{equation} \label{KMT23mf}
    K_{\rm mf}=\sum_{m=0}^{M-1} (M-m+1) s^{(m)}_n+i_n-\ln \frac{s}{s_0}= \sum_{m=0}^{M-1} (M-m+1) s^{(m)}_0+i_0.
\end{equation}
In full analogy with the cases considered above, in the mean-field limit, the value of this constant of motion is related to that of Eq.~(\ref{KMT23}) as $K=NK_{\rm mf}$.

\section{Approximate constants of motion in SIR epidemics} \label{SIR}

The SIR (susceptible-infected-recovered) model for epidemic propagation is among the most fundamental mathematical representations of the spreading of a contagious disease in a population of interacting agents. It belongs to a wide class of models where contact processes are used to describe the elementary events intervening in disease dissemination \cite{epi1,epi2,epi3}. In the SIR  model, the contact of an infected agent and a susceptible agent may lead the latter to become infected. In contrast with the Maki-Thompson model studied above, however, the transition from the infected state to the recovered state occurs spontaneously, without intervention of other agents. Recovered agents, in turn, cannot change their state.  

Although the SIR model is frequently formulated in terms of mean field equations \cite{epi3}, it also admits several versions in the form of a stochastic contact process. By analogy with the Maki-Thompson process, at each time step, we here choose an infected agent at random. With probability $u$, the agent becomes recovered. With the complementary probability, $1-u$, another agent is chosen at random from the whole population and, if this second agent is susceptible, it becomes infected. Table \ref{tab4} shows the elementary events of the model, their probabilities, and the corresponding constants of motion.

\begin{table}[H] 
\caption{As in Table \ref{tab1} for the SIR epidemiological model. \label{tab4}}
\begin{tabularx}{\textwidth}{lll}
\toprule
\textbf{event}	& \textbf{probability}	 &\textbf{constants of motion}  \\
\midrule
${\rm I} \to {\rm R}$& $u$ &  $S_n$, $I_n+n$, $R_n-n$\\
${\rm I}+{\rm S} \to {\rm I}+{\rm I}$ & $(1-u)S_n/(N-1)$ &  $S_n+n$, $I_n-n$, $R_n$ \\
${\rm I}+\{ {\rm I},{\rm R} \} \to {\rm I}+\{ {\rm I},{\rm R} \}$ &  $(1-u)[1-S_n/(N-1)]$ & $S_n$, $I_n$,  $R_n$\\
\bottomrule
\end{tabularx}
\end{table}

Proceeding as for the Maki-Thompson model to find joint conserved quantities for the stochastic process, we note that --since, now, the process is defined  by three individual events-- the coefficients in a linear combination of the individual constants of motion  must satisfy two conditions,  
\begin{equation} \label{cond2}
    \alpha-\beta=\beta-\gamma=0,
\end{equation}
instead of the only condition of Eq.~(\ref{cond1}). Consequently, there is only one independent solution, $\alpha=\beta=\gamma=1$, corresponding to the trivial constant of motion $N=S_n+I_n+R_n$. Thus, the SIR stochastic process has no other conserved quantity.

We note, however, that the first event in  Table \ref{tab4}, on the one hand, and the second and third events, on the other, have their own constants of motion, respectively,
\begin{equation}
    K_1=S_n+I_n+n, \ \ \ \ \ K_2=S_n+I_n.
\end{equation}
This suggests that a combination of $K_1$ and $K_2$, weighted with the total probabilities of the corresponding events,  
\begin{equation} \label{Ktilde}
    \widetilde K_n = uK_1+(1-u)K_2= S_n+I_n+u n ,
\end{equation}
may play a role similar to that of a constant of motion, although not being an exactly conserved quantity itself. To assess this conjecture, we explicitly calculate the stochastic evolution of $\widetilde K_n$, finding
\begin{equation} \label{KtildeEVOL}
     \widetilde K_{n+1} = 
     \begin{cases}
         \widetilde K_n-(1-u), & \mbox{with probability $u$,} \\ 
         \widetilde K_n+u, & \mbox{with probability $1-u$.}
     \end{cases}
\end{equation}
This stochastic process represents a one-dimensional symmetric random walk: its mean value and variance evolve  as
\begin{equation} \label{sigma}
    \langle \widetilde K\rangle_{n+1} = \langle \widetilde K\rangle_n, \ \ \ \ \ \sigma^2_{n+1} = \sigma^2_n + u(1-u),
\end{equation}
respectively. In other words, $\widetilde K_n$ is an {\em approximate} constant of motion, in the sense that its mean value is a conserved quantity. Its standard deviation, in turn, grows as $n^{1/2}$, like in any ordinary diffusion process. Note that, in combination with the trivial constant of motion $N$, $\widetilde K_n$ straightforwardly yields the mean number of recovered agents over realizations with initial condition $R_0$:
\begin{equation}
    \langle R\rangle_n = R_0+u n.
\end{equation}

The approximate constant of motion  $\widetilde K_n$ acquires relevance when considering the mean-field version of the SIR model. To show this, first, we use the master equation of the process,
\begin{equation}
    P_{n+1} (S,I) =u P_n(S,I+1)+ (1-u)\frac{S+1}{N-1} P_n(S+1,I-1)+\left[1-(1-u)\frac{S}{N-1} \right] P_n(S,I),
\end{equation}
to find evolution equations for the mean numbers $\langle S \rangle_n$ and $\langle I \rangle_n$:
\begin{equation} \label{meanSI}
    \langle S \rangle_{n+1} = \left( 1-\frac{1-u}{N-1}\right)\langle S \rangle_n, \ \ \ \ \ 
    \langle I \rangle_{n+1}= \langle I \rangle_n+(1-u) \frac{\langle S \rangle_n}{N-1}-u. 
\end{equation}
The solution to the first of these equations is
\begin{equation} \label{sol44}
    \langle S \rangle_n = \left( 1-\frac{1-u}{N-1}\right)^n  S_{0}
\end{equation}
for any fixed initial condition $S_0$. Next, from Eqs.~(\ref{meanSI}) and operating as for the Maki-Thompson model, we get the mean-field description:
\begin{equation}
    \frac{\text d s}{\text d t} = -\frac{(1-u)si}{\tau}, \ \ \ \ \ \frac{\text d i}{\text d t}= \frac{(1-u)si}{\tau} -\frac{ui}{\tau}, \ \ \ \ \ \frac{\text d r}{\text d t}= \frac{ui}{\tau}.
\end{equation}
Now, calculating ${\text d i}/{\text d s}$ and integrating with respect to $s$, we find a mean-field constant of motion:
\begin{equation}
    K_{\rm mf} = s+i - \frac{u}{1-u} \ln \frac{s}{s_0}=s_0+i_0.
\end{equation}
Much as in Eq.~(\ref{equiv}), in the mean-field limit we can write
\begin{equation}
    \ln \frac{s(t)}{s_0} = \ln \frac{ \langle S \rangle_n}{S_0} = -\frac{(1-u) n}{N},
\end{equation}
where we have used Eq.~(\ref{sol44}). Comparing Eq.~(\ref{Ktilde}), this shows that, in the limit, the {\em approximate} constant of motion $\widetilde K_n$ coincides with the {\em exact} mean-field conserved quantity $K_{\rm mf}$ up to a multiplicative constant: $\widetilde K_n = N K_{\rm mf}$.

\begin{figure}[H]
\includegraphics[width=10 cm]{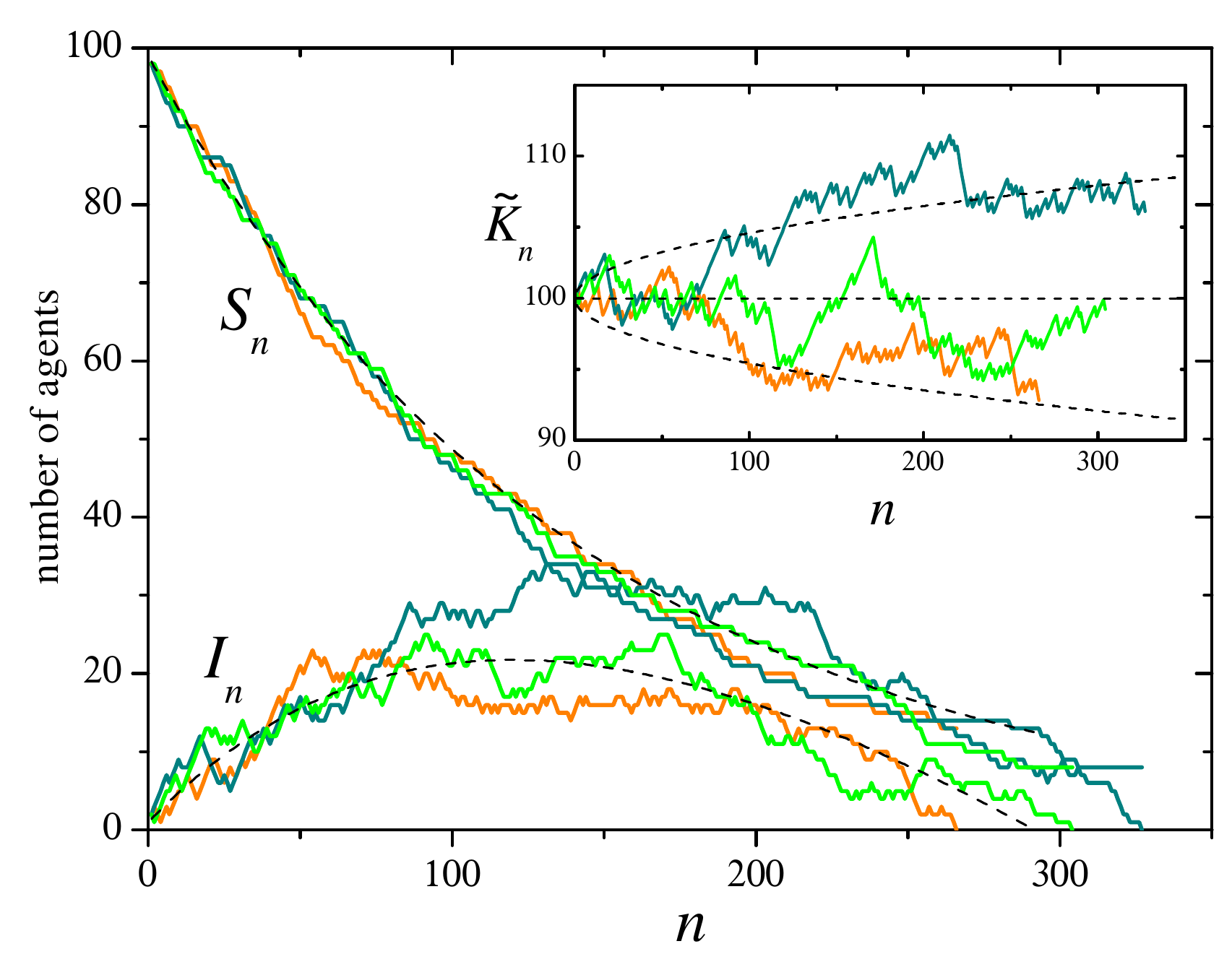}
\caption{Three realizations of the SIR stochastic process defined by the events of Table \ref{tab4}, with $N=100$, $u=0.3$, and initial conditions $S_0=N-1$ and $I_0=1$. In the main panel, curves show the evolution of the number of susceptible and infected agents, $S_n$ and $I_n$, as a function of the evolution step $n$. Each color correspond to a single realization. Note that each realization ends when the number of infected agents reaches zero. Dashed curves represent the solutions of Eqs.~(\ref{meanSI}) for the mean values $\langle S \rangle_n$ and $\langle I \rangle_n$. The inset shows the evolution of the approximate constant of motion $\widetilde K_n$, Eq.~(\ref{Ktilde}), for the same realizations. The horizontal dashed line is the expected mean value,  $\langle \widetilde K \rangle_n=S_0+I_0$, and the dashed curves indicate the corresponding standard deviation, $\sigma_n$, given by Eq.~(\ref{sigma}). \label{fig01}}
\end{figure}

The main panel of Fig.~\ref{fig01} illustrates the evolution of the number of susceptible and infected agents along three realizations of the stochastic process, for a system with $N=100$ agents and recovery probability $u=0.3$, from an initial condition with one infected agent and no refractory agents. A comparison with the mean evolution given by Eqs.~(\ref{meanSI}) is provided. The inset shows the evolution of the approximate constant of motion $\widetilde K_n$, together with its expected mean value and standard deviation.

\subsection{SIR epidemics on random networks}

An important extension of epidemiological models based on contact processes, aimed at approaching a more realistic representation of social structure, consists in considering populations distributed on a network \cite{en1,en2,en3}. In this variant, agents are situated at the network nodes, and network links represent their mutual contacts, so that infection can only occur between neighbor agents. The heterogeneity induced by the network of contacts on the population structure implies that susceptible and infected agents have more chances of, respectively, becoming infected and transmitting the infection when their number of neighbors --namely, their degree-- is larger. The mathematical description of the process thus requires discerning between agents with different degrees. For each possible degree $k$, we denote by ${\rm S}^{(k)}$, ${\rm I}^{(k)}$, and ${\rm R}^{(k)}$, the corresponding number of agents in each state of the SIR epidemics.

The SIR stochastic contact process on networks is here implemented as follows. At each evolution step, we choose at random a network link connected to (at least) one infected agent. With probability $u$, this infected agent becomes recovered. With the complementary probability, $1-u$, if the second agent connected to the chosen link is susceptible, it becomes infected. Clearly, since one link is chosen at each step, the probability that any given infected or susceptible agent becomes involved in an event now depends on the respective degree $k$. The second column of Table \ref{tab5} quotes the probabilities for each event in the case of a network with links distributed at random and with no statistical correlation between the degrees of neighbor agents.  

\begin{table}[H] 
\small 
\caption{Events in the SIR epidemiological model evolving on a random, degree-uncorrelated network. The first row holds for every value of the degree $k$. The second and third rows hold for every pair $k$, $k'$. In the third row, $\{ {\rm I}^{(k')},{\rm R}^{(k')} \}$ stands for either ${\rm I}^{(k')}$ or ${\rm R}^{(k')}$, and $N^{(k')}$ is the total number of agents with degree $k'$.  Moreover,  $\vartheta_n = \sum_k k I_n^{(k)}$ and $\zeta_n= \vartheta_n z (N-1)$ are normalization constants, with $z$ the mean number of neighbors per site. In the first two rows, for brevity, the third column only shows the constants of motion which combine numbers of agents and the index of the evolution step $n$. All the other numbers of agents are also constants of motion for both events.  \label{tab5}}
\begin{tabularx}{\textwidth}{lll}
\toprule
\textbf{event}	& \textbf{probability}	 &\textbf{constants of motion}  \\
\midrule
${\rm I}^{(k)} \to {\rm R}^{(k)}$& $ukI_n^{(k)}/\vartheta_n$ &  $I_n^{(k)}+n$, $R_n^{(k)}-n$\\
${\rm I}^{(k)}+{\rm S}^{(k')} \to {\rm I}^{(k)}+{\rm I}^{(k')}$ & $(1-u)kk'I_n^{(k)} S_n^{(k')}/\zeta_n$ &  $S_n^{(k')}+n$, $I_n^{(k)}-n$ \\
${\rm I}^{(k)}+\{ {\rm I}^{(k')},{\rm R}^{(k')} \} \to {\rm I}^{(k)}+\{ {\rm I}^{(k')},{\rm R}^{(k')} \}$ & $(1-u)kk'I_n^{(k)} [N^{(k')}-S_n^{(k')}]/\zeta_n$ & $S_n^{(k)}$, $I_n^{(k)}$,  $R_n^{(k)} \ \ \ \forall k$\\
\bottomrule
\end{tabularx}
\end{table}

Linear combinations of the constants of motion in the third column of Table \ref{tab5} show that the only independent conserved quantities common to all the events are, trivially, the total number of agents with each degree $k$, $N^{(k)}=S_n^{(k)}+I_n^{(k)}+R_n^{(k)}$, for $k=1,2, \dots$. Meanwhile, a combination analogous to $\widetilde K_n$ in Eq.~(\ref{Ktilde}), namely,
\begin{equation}
    \widetilde K_n^{(0)} =S_n+I_n+un = \sum_k \left[ S_n^{(k)}+ I_n^{(k)} \right]+un,
\end{equation}
where $S_n$ and $I_n$ are the total numbers of susceptible and infected agents at step $n$, satisfies the same stochastic evolution equation as $\widetilde K_n$ in Eq.~(\ref{KtildeEVOL}), and is thus an approximate constant of motion.

It turns out, however, that $\widetilde K_n^{(0)}$ is just but one among infinitely many combinations that act as approximate constants of motion for the SIR model on networks. To see this, consider for example the quantities $Q_n^{(\gamma)}= \sum_k (k/z)^\gamma [ S_n^{(k)}+ I_n^{(k)} ] $, for any value of the exponent $\gamma$, where $z$ is the mean number of neighbors per network site. In an event as in the first line of Table \ref{tab5}, where an infected agent becomes recovered at step $n$, we have $Q_{n+1}^{(\gamma)}=Q_n^{(\gamma)}-(k_n/z)^\gamma$, where $k_n$ is the degree of the infected agent involved in that event. On the other hand, for the events in the second and third rows, we have $Q_{n+1}^{(\gamma)}=Q_n^{(\gamma)}$. Averaging over the possible values of $k_n$ with their corresponding probabilities (second column of Table \ref{tab5}), it turns out that the quantity
\begin{equation} \label{Kg}
    \widetilde K_n^{(\gamma)}=Q_n^{(\gamma)}+u \sum_{m=0}^{n-1} \Theta_m^{(\gamma)} = \sum_k \left( \frac{k}{z}\right)^\gamma \left[ S_n^{(k)}+ I_n^{(k)} \right]+u \sum_{m=0}^{n-1} \Theta_m^{(\gamma)} , 
\end{equation}
with
\begin{equation} \label{Theta}
    \Theta_m^{(\gamma)}=\frac{\sum_k k^{\gamma+1} I_m^{(k)}}{z^\gamma\sum_k k I_m^{(k)}},
\end{equation}
is driven by the stochastic process\begin{equation} \label{KtildegEVOL}
     \widetilde K_{n+1}^{(\gamma)} = 
     \begin{cases}
         \widetilde K_n^{(\gamma)}-(1-u) \Theta_n^{(\gamma)} , & \mbox{with probability $u$,} \\ 
         \widetilde K_n^{(\gamma)}+u \Theta_n^{(\gamma)}, & \mbox{with probability $1-u$;}
     \end{cases}
\end{equation}
cf.~Eq.~(\ref{KtildeEVOL}). For this process,
\begin{equation} \label{sigmag}
\langle \widetilde K^{(\gamma)}\rangle_{n+1} = \langle \widetilde K^{(\gamma)}\rangle_n, \ \ \ \ \ \sigma^2_{n+1} = \sigma^2_n + u(1-u)\left[ \Theta_n^{(\gamma)} \right]^2,
\end{equation}
give the evolution of the mean value and the standard deviation of $\widetilde K^{(\gamma)}_n$; cf.~Eq.~(\ref{sigma}). In other words, $\widetilde K^{(\gamma)}_n$ is an approximate constant of motion for any value of $\gamma$. Linear combinations of  $\widetilde K^{(\gamma)}_n$ for different values of $\gamma$ would produce infinitely many other approximate conserved quantities.

Mean-field equations for the stochastic SIR model on networks are obtained by assigning to each event a duration which appropriately takes into account the probability that an infected agent of degree $k$ becomes involved in that event. Equation (\ref{tau}) now becomes
\begin{equation}
    t_{n+1} -t_{n} = \frac{\tau k_n}{\sum_{k'} k'I_n^{(k')} }\equiv   \frac{\tau k_n}{\vartheta_n},
\end{equation}
with $\vartheta_n$ defined as in the caption of Table \ref{tab5}. Here, $k_n$ is the degree of the infected agent involved at step $n$ and, as in Eq.~(\ref{tau}), $\tau$ fixes time units. The resulting equations for the fractions $s^{(k)}= \langle S^{(k)} \rangle/N$, $i^{(k)}= \langle I^{(k)} \rangle/N$, and $r^{(k)}= \langle R^{(k)} \rangle/N$  are
\begin{equation} \label{43}
    \begin{split}
        \frac{\text d s}{\text dt }^{(k)} &= - \frac{(1-u)k s^{(k)} }{z\tau} \sum_{k'} \frac{k'}{z} i^{(k')}, \\
        \frac{\text d i}{\text dt }^{(k)} &= \frac{(1-u)k s^{(k)} }{z\tau} \sum_{k'} \frac{k'}{z} i^{(k')}-\frac{ui^{(k)}}{\tau} ,  \\
        \frac{\text d r}{\text dt }^{(k)} &= \frac{ui^{(k)}}{\tau} , \\
    \end{split}
\end{equation}
for each degree $k$. Operating on these equations, it can be straightforwardly shown that the quantity
\begin{equation} \label{Kmf5}
    K_{\rm mf} = \sum_{k'} \frac{k'}{z} \left[s^{(k')}+i^{(k')} \right] -\frac{u}{1-u} \frac{z}{k} \ln \frac{s^{(k)}}{s^{(k)}_0}
\end{equation}
is a constant of motion, with $s^{(k)}_0$ the initial density of susceptible agents with $k$ neighbors. We stress that, in spite of its functional form, the last term in the right-hand side of Eq.~(\ref{Kmf5}) is independent of $k$, as it becomes clear if the first of Eqs.~(\ref{43}) is rewritten as
\begin{equation}
\frac{1}{k} \frac{\text d }{\text dt }\ln s ^{(k)} = - \frac{1-u }{z\tau} \sum_{k'} \frac{k' i^{(k')}}{z} .
\end{equation}

Comparing Eqs.~(\ref{Kmf5}) and (\ref{Kg}), we realize that the approximate constant of motion of the stochastic process to be related to  $ K_{\rm mf}$ is $\widetilde K^{(1)}_n$. Up to a factor $N$, the summations in both constants are the same, so that the comparison must be carried out between the last term in the right-hand side of each equation. In the stochastic process, the mean evolution of susceptible agents is 
\begin{equation} \label{lt44}
    \langle S^{(k)}\rangle_n =\left[1-\frac{(1-u)k}{z(N-1)} \right]^n S^{(k)}_0.
\end{equation}
Replacing in the last term  of Eq.~(\ref{Kmf5}), we get
\begin{equation}
    -\frac{u}{1-u} \frac{z}{k} \ln \frac{s^{(k)}}{s^{(k)}_0} = \frac{un}{N},
\end{equation}
in the large-$N$ limit. As for the last term of Eq.~(\ref{Kg}) with $\gamma=1$, $u\sum_m \Theta_m^{(1)}$, Eq.~(\ref{Theta}) implies that, if infected agents are evenly distributed over the network, we can approximate $\Theta_m^{(1)} =\langle k^2 \rangle /z^2$ for all $m$. Within this approximation, which should improve as the system size grows, we have
\begin{equation}
    u\sum_{m=0}^{n-1} \Theta_m^{(1)} = un \frac{\langle k^2 \rangle}{z^2}.
\end{equation}
Taking into account Eq.~(\ref{lt44}), it is clear that the expected proportionality $\widetilde K^{(1)}_n=N K_{\rm mf}$ only holds if $\langle k^2 \rangle=z^2$, i.~e.~for a {\em regular} random network \cite{newman}, where all agents have the same number of neighbors. Note that this condition is verified, in particular, for a fully connected network, where $k=N-1$ for all agents. In this case, all agents can interact with each other, and we recover the SIR epidemic model without network.

\begin{figure}[H]
\centering
\includegraphics[width=10 cm]{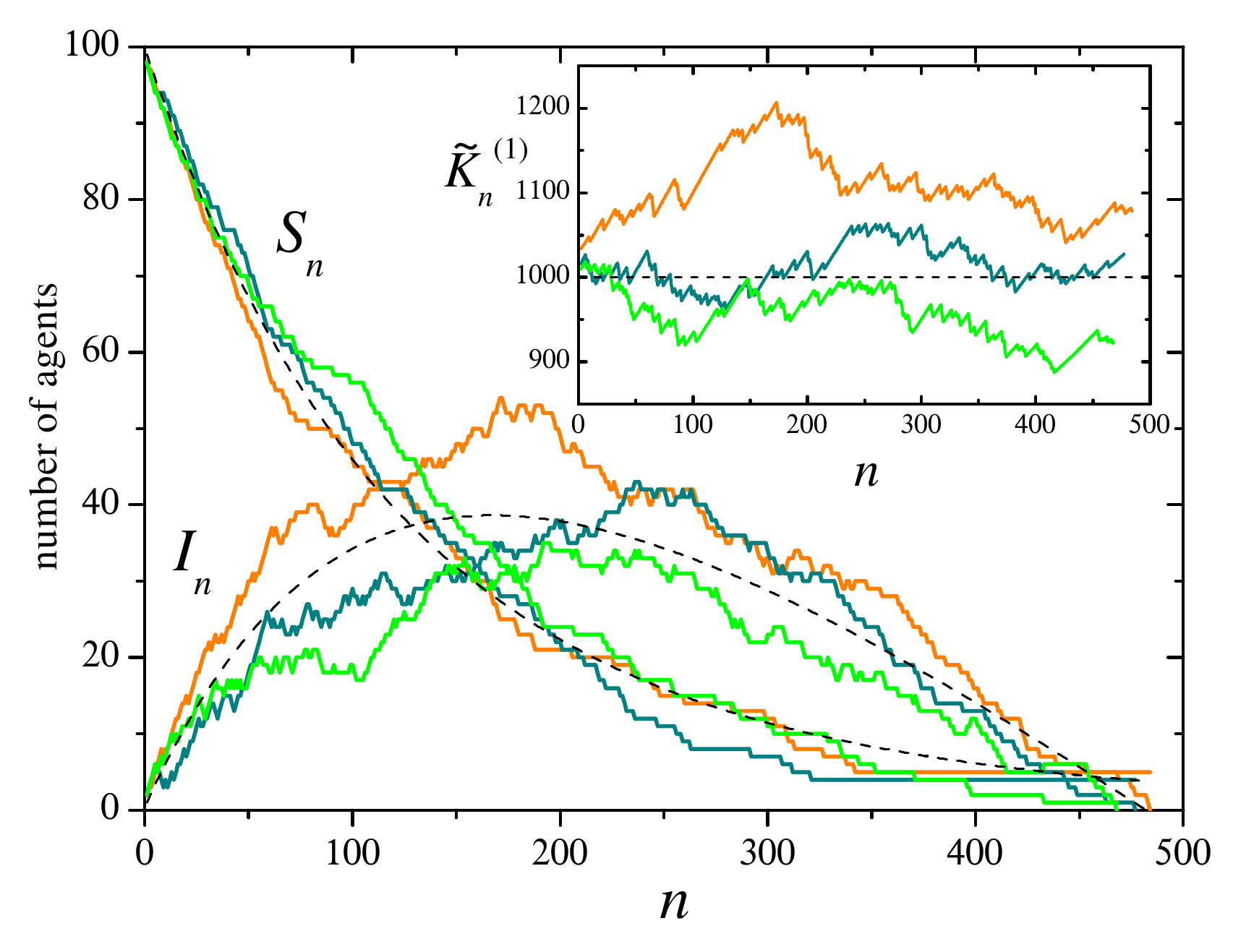}
\caption{Three realizations of the SIR stochastic process defined by the events of Table \ref{tab5}, with $N=100$, $u=0.2$, and initial conditions $S_0=N-1$ and $I_0=1$. The system evolves on an Erd\H{o}s-R\'enyi random network, different for each realization, with an average of $z=10$ neighbors per site. In the main panel, curves show the evolution of the total number of susceptible and infected agents, $S_n$ and $I_n$, as a function of the evolution step $n$. Each color correspond to a single realization. Dashed curves represent the mean values of $S_n$ and $I_n$ averaged over realizations of the evolution and the underlying network.  for the same realizations, the inset shows the evolution of the approximate constant of motion $\widetilde K_n^{(1)}$, given by Eq.~(\ref{Kg}) with $\gamma=1$. The horizontal dashed line is the expected mean value,  $\langle \widetilde K^{(1)} \rangle_n=zN$. \label{fig02}}
\end{figure}

Figure \ref{fig02} shows, in the main panel, results for three realizations of the SIR stochastic process on Erd\H{o}s-R\'enyi random networks \cite{newman} with $N=100$ agents, and recovery probability $u=0.2$, from an initial condition with one infected agent and no refractory agents. On the average, each agent has $z=10$ neighbors. Full curves stand for the evolution of the total number of susceptible and infected agents, respectively, $S_n$ and $I_n$. Dashed curves correspond to their expected values, averaged over realizations of both the evolution and the underlying network. The inset shows the evolution of the approximate constant of motion $\widetilde K_n^{(1)}$, given by Eq.~(\ref{Kg}) for $\gamma=1$. The horizontal dashed line indicates its expected mean value, $zN$.

\section{Nonlinear and ``failed'' constants of motion}

The conserved quantities derived for the stochastic processes discussed in the preceding sections were built as linear combinations of the constants of motion associated to each event, which we have quoted in Tables \ref{tab1} to  \ref{tab5}. Nothing precludes, however, that a suitably chosen {\em nonlinear} combination of the individual constants of motion can fulfill that role for the whole process. To illustrate this situation, we consider a variant of the Maki-Thompson model of rumor propagation analyzed in Section \ref{MTrp}, where the contact between two infected agents results in {\em both} agents becoming refractory. Any other event occurs as in the original model. Table \ref{tab6} lists the events, their probabilities, and the individual constants of motion for this variant.

\begin{table}[H] 
\caption{As in Table \ref{tab1}, for a variant of  the Maki-Thompson model of rumor propagation where the contact of two infected agents results in both agents becoming refractory. 
\label{tab6}}
\begin{tabularx}{\textwidth}{lll}
\toprule
\textbf{event}	& \textbf{probability}	 &\textbf{constants of motion}  \\
\midrule
${\rm I}+{\rm S} \to {\rm I}+{\rm I}$ & $S_n/(N-1)$  & $S_n+n$, $I_n-n$, $R_n$ \\
${\rm I}+{\rm I} \to {\rm R}+{\rm R}$ &  $(I_n-1)/(N-1)$ &  $S_n$, $I_n+2n$, $R_n-2n$\\
${\rm I}+ {\rm R}  \to {\rm R}+ {\rm R}  $ &  $R_n/(N-1)$ &  $S_n$, $I_n+n$, $R_n-n$\\
\bottomrule
\end{tabularx}
\end{table}

The only linear combination of the constants of motion in Table \ref{tab6} which acts as a conserved quantity for the three events is, not unexpectedly, the total number of agents $N=S_n+I_n+R_n$. In principle, thus, we cannot rely on other combinations to get information on the evolution of the stochastic process. However, some algebra on the mean-field equations,
\begin{equation}  \label{MTMF3}
 \frac{\text d s}{\text d t}= -\frac{si}{\tau}, \ \ \ \ \  \frac{\text d i}{\text d t}=  \frac{si -i(2i+r)}{\tau}, \ \ \ \ \ \frac{\text d r}{\text d t}= \frac{i(2i+r)}{\tau},
\end{equation}
shows that 
\begin{equation} \label{kmfMT4}
    K_{\rm mf} = \frac{i+1}{s}+2\ln \frac{s}{s_0} \equiv \frac{i_0+1}{s_0}
\end{equation}
is a constant of motion along the mean-field dynamics. This suggests that a constant of motion for the stochastic process, now involving a nonlinear combination of the number of agents in each state, may in fact exist.  

To explore this possibility, we first write down the master equation for the probability $P_n(S,I)$ that, at step $n$, the numbers of susceptible and infected agents are $S$ and $I$, respectively: 
\begin{equation}
P_{n+1} (S,I) =\frac{S+1}{N-1}P_n(S+1,I-1)+\frac{I+1}{N-1} P_n(S,I+2)+\left( 1-\frac{S+I}{N-1} \right) P_n(S,I+1),
\end{equation}
where each term in the right-hand side represents the contribution of one of the events in Table \ref{tab6}. From this master equation, we reobtain Eq.~(\ref{deltaS}) for the evolution of $\langle S \rangle_n$. For $\langle I \rangle_n$, in turn, we get
\begin{equation} \label{In6}
    \langle I \rangle_{n+1} =  \langle I \rangle_{n} + \frac{1}{N-1} \left[ \langle S \rangle_{n} -2 ( \langle I \rangle_{n}-1) - \langle R \rangle_{n} \right],
\end{equation}
with $\langle R \rangle_{n}=N-\langle S \rangle_{n}-\langle I \rangle_{n}$. Taking into account the evolution laws for $\langle S \rangle_{n}$ and $\langle I \rangle_{n}$, Eqs.~(\ref{In6}) and (\ref{deltaS}), it turns out that the combination
\begin{equation} \label{49v}
    \widetilde K = \frac{\langle I\rangle_n+N-2}{\langle S \rangle_n}-\frac{2}{N-2} n
\end{equation}
is preserved along the evolution. Note however that the nature of this constant of motion is different from that of the examples considered in the preceding sections. In fact, it does not coincide with either a combination of the variables $S_n$ and $I_n$ or the mean value of any such combination. Rather, it is a nonlinear combination of mean values. In any case, in view of the solution for $\langle S\rangle_n$ given by Eq.~(\ref{Sn1}), in the limit of large $N$, $\widetilde K$ coincides with the mean-field constant of motion $K_{\rm mf}$ given by Eq.~(\ref{kmfMT4}). Moreover, using the solution for $\langle S\rangle_n$ and the form of $\widetilde K$, it is possible to find the solution for the mean number of infected agents, which reads
\begin{equation}
    \langle I\rangle_n=\left(I_0+N-2 +\frac{2n}{N-2}  S_0 \right) \left( 1-\frac{1}{N-1}\right)^n-N+2 .
\end{equation}

Finally, we consider a complementary variant of the Maki-Thompson process, where an infected agent must necessarily contact a refractory agent to become in turn refractory. Namely, the contact between two infected agents produces no change. Table \ref{tab7} shows the relevant parameters. As in the preceding variant, the only linear constant of motion is the total number of agents, $N=S_n+I_n+R_n$. 

\begin{table}[H] 
\caption{As in Table \ref{tab1}, for a variant of  the Maki-Thompson model of rumor propagation where the contact of two infected agents produces no change. 
\label{tab7}}
\begin{tabularx}{\textwidth}{lll}
\toprule
\textbf{event}	& \textbf{probability}	 &\textbf{constants of motion}  \\
\midrule
${\rm I}+{\rm S} \to {\rm I}+{\rm I}$ & $S_n/(N-1)$  & $S_n+n$, $I_n-n$, $R_n$ \\
${\rm I}+{\rm I} \to {\rm I}+{\rm I}$ &  $(I_n-1)/(N-1)$ &  $S_n$, $I_n$, $R_n$\\
${\rm I}+ {\rm R}  \to {\rm R}+ {\rm R}  $ &  $R_n/(N-1)$ &  $S_n$, $I_n+n$, $R_n-n$\\
\bottomrule
\end{tabularx}
\end{table}

For this process, the mean-field equations read 
\begin{equation}  \label{MTMF4}
 \frac{\text d s}{\text d t}= -\frac{si}{\tau}, \ \ \ \ \  \frac{\text d i}{\text d t}=  \frac{(
 s-r)i }{\tau}, \ \ \ \ \ \frac{\text d r}{\text d t}= \frac{ri}{\tau}.
\end{equation}
Taking the ratio of the first and third of these equations, it becomes apparent that the product
\begin{equation} \label{kmfMT5}
    K_{\rm mf} = sr=s_0r_0
\end{equation}
is a conserved quantity. In order to relate $K_{\rm mf}$ to a constant of motion for the stochastic process, it is convenient to use the master equation for the probability $P_n(S,R)$: 
\begin{equation} \label{ME7}
P_{n+1} (S,R) =\frac{S+1}{N-1}P_n(S+1,R)+\frac{R-1}{N-1} P_n(S,R-1)+\left( 1-\frac{S+R}{N-1} \right)P_n(S,R),
\end{equation}
which yields
\begin{equation}
     \langle SR\rangle_{n+1}=\langle SR\rangle_n .
\end{equation}
In other words, the product $\widetilde K_n= S_n R_n$ is an approximate conserved quantity for the stochastic process, whose mean value is proportional to the constant of motion of the mean-field equations: $\langle \widetilde K \rangle_n = N^2 K_{\rm mf}$.

As a final warning, we remark that the mean number of susceptible and refractory agents here evolve as
\begin{equation}
    \langle S\rangle_n = \left(1-\frac{1}{N-1}\right)^n S_0, \ \ \ \ \ \langle R\rangle_n = \left(1+\frac{1}{N-1}\right)^n R_0,
\end{equation}
from initial conditions $S_0$ and $R_0$. If, operating as for the preceding variant in Eq.~(\ref{49v}), we naively compute the product
\begin{equation}
    \langle S\rangle_n \langle R\rangle_n =\left[ 1-\frac{1}{(N-1)^2}\right]^n S_0 R_0,
\end{equation}
we get a quantity that is {\em not} a constant of motion, but {\em decreases monotonically} as $n$ grows. In other words, in the present case, a combination of stochastic variables inspired by a conserved quantity at the mean-field level, {\em fails} to give origin to a constant of motion. Only in the limit $N\to \infty$ does the procedure work, yielding a conserved quantity which, as expected, is proportional to that of the mean-field equations.   

\section{Conclusion}

In this paper, we have considered a variety of stochastic contact processes where agents in a population change their individual state through random events of pair interaction. The focus has been put on the existence of conserved quantities, and on the role of these constants of motion in the mathematical analysis of such processes. Our procedure can be summarized as follows. For each elementary interaction event in the process, we first identify its individual constants of motion, which immediately derive from the changes in the  state of the involved agents. Then, we find combinations of the individual constants of motion that are conserved quantities of all the interaction events and, therefore, are strictly conserved along the whole dynamics. In principle, these combinations could be arbitrary functions of the individual constants of motion, but we have mainly considered linear combinations, for which the procedure can be carried out systematically. Individual constants of motion can also be weighted by the probabilities of the corresponding events, and then linearly combined to get approximately conserved quantities, which fluctuate randomly around a well defined value as the system evolves. Exact and approximate constants of motion can finally be used to ease the analytical treatment by reducing the number of relevant independent variables. Comparison with their counterparts in a mean-field description of the dynamics also proves to be meaningful: in the appropriate limit, the values of stochastic and mean-field conserved quantities usually coincide with each other. 

For the Maki-Thompson model of rumor propagation, which is defined by two possible interaction events with three accessible states, the above procedure yields two exact constants of motion. The first, trivial one is just the total number of agents. Together with the second constant of motion, it reduces the state space of the system to one single dimension which, using the master equation of the stochastic process, yields the full solution to the problem. Moreover, in the continuous-time limit for large systems, the values of the two constants of motion coincide with those found for the corresponding mean-field equations, even when their functional forms are not the same. Extensions of the Maki-Thompson model, adding intermediate states to the evolution of each agent, also contain two exact constants of motion. However, since the state space in now higher-dimensional, the usefulness of the conserved quantities in solving the problem is more limited.   

In the SIR model of disease spreading, where agents can access three different states as well, the number of possible elementary events grows to three and, consequently, there is only one exact constant of motion, namely, the number of agents. However, it is still possible to find an approximate nontrivial conserved quantity. Its dynamics coincide with those of a symmetric random walker, so that its mean value remains constant while its standard deviation grows as the square root of the number of steps. Thus, this approximate constant of motion plays the same role, with respect to the evolution averaged over realizations, as the exact nontrivial conserved quantity of the Maki-Thompson model, and allows for the full solution to the stochastic problem in terms of the mean values of the relevant variables. Moreover, as in the case of rumor propagation, its value coincides with a constant of motion of the mean-field equations in the appropriate limit. Similar results are obtained for the SIR model evolving on random networks, although in this case the exact coincidence with the mean-field constant of motion is restricted to the case of regular networks. 

Finally, we have considered a couple of stochastic processes in which the corresponding mean-field equations suggest the existence of conserved quantities given by nonlinear combinations of the constants of motion of the elementary events. In these cases, it was possible to show that a suitable extension of the mean-field conserved quantities to the level of the stochastic description does yield approximate constants of motion, although not all extensions successfully connect conserved quantities at the two levels.

Although the models for rumor and disease propagation considered in this paper provide a fair illustration of how conserved quantities in stochastic process yield useful information on their evolution, and give insight on the connection with mean-field descriptions, it is clear that these examples do not exhaust the possible uses of such techniques. Stochastic models for chemical reactions \cite{vanKampen,ch1}, biological population dynamics \cite{eco1,eco2,eco3,eco4}, and socioeconomic phenomena \cite{Liggett,soc} are foreseeable fields of application of the same approach.


\end{document}